\documentclass[10pt,showpacs,twocolumn,nofootinbib,aps,prd]{revtex4-1}
\usepackage{CJK}
\usepackage{amsmath,amssymb,bm,mathrsfs}
\usepackage{mathtools,mathptmx,array,slashed}
\usepackage{graphicx,graphics,subfigure}
\usepackage{ctable,multirow}
\usepackage{tabularx}
\usepackage[colorlinks=true,linkcolor=blue,citecolor=blue,urlcolor=blue]{hyperref}

\newcommand{\meq}[1]{(\ref{#1})}

\newcommand{\pp}{\partial}

\def\be{\begin{equation}}
\def\ee{\end{equation}}
\def\bea{\begin{eqnarray}}
\def\eea{\end{eqnarray}}
\def\ba{\begin{aligned}}
\def\ea{\end{aligned}}

\def\pp{\partial}

\begin{document}

%\begin{CJK*}{GBK}{song}

\title{Light rings and shadows of static black holes in effective quantum gravity II: A new solution without Cauchy horizons}

\author{Wentao Liu$^{1}$}
\email{wentaoliu@hunnu.edu.cn}

\author{Di Wu$^{2}$}
\email{Corresponding author: wdcwnu@163.com}
% https://orcid.org/0000-0002-2509-6729

\author{Jieci Wang$^{1}$}
\email{Corresponding author: jcwang@hunnu.edu.cn}

\affiliation{$^{1}$Department of Physics, Key Laboratory of Low Dimensional Quantum Structures and Quantum Control of Ministry of Education, and Synergetic Innovation Center for Quantum Effects and Applications, Hunan Normal University, Changsha, Hunan 410081, People's Republic of China \\
$^{2}$School of Physics and Astronomy, China West Normal University, Nanchong, Sichuan 637002, People's Republic of China}

\date{\today}

\begin{abstract}
Among the three known types of static solutions proposed within the Hamiltonian constraint approach to effective quantum gravity (EQG), the first two have been extensively investigated, whereas the third type-which preserves general covariance, is free of Cauchy horizons, and was only recently obtained-remains relatively unexplored.
This solution can describe a black hole with an event horizon for certain parameter ranges, or a horizonless compact object beyond those ranges.
In this paper, we focus on the third type and show that its light rings feature both stable and unstable branches, and that the black hole shadow size grows with the quantum parameter-unlike in the first two types.
However, when we account for both the shadow and the lensing ring, the overall behavior closely resembles that of the second type, in which an increasing quantum parameter leads to a larger portion of the lensing ring being occupied by the shadow.
This feature can serve as a hallmark of black holes in EQG, offering a potential way to distinguish them from their GR counterparts.
Remarkably, the parameter ranges under which the solution remains a black hole are highly consistent with the current observational constraints on black hole shadows, lending strong support to the classification of the third type of compact object in EQG as a black hole endowed with an event horizon.
\end{abstract}
\maketitle

%\end{CJK*}

%%%%%%%%%%%%%%%%%%%%%%%%%%%
\section{Introduction}
%%%%%%%%%%%%%%%%%%%%%%%%%%%
The existence of singularities and their inconsistency with quantum physics suggest that general relativity (GR) might not be the ultimate theory of spacetime \cite{PRL14-57,PRL130-101501,Calza:2024xdh}.
In the realm of loop quantum gravity, research on black holes aims to unify quantum mechanics with general relativity, addressing challenges such as black hole singularities and the information paradox, while also probing the quantum nature of spacetime.
A recent study \cite{2407.10168} introduces the covariance equation within Effective Quantum Gravity (EQG).
By solving this equation, Zhang \textit{et al}. derived three distinct types of Hamiltonian constraints \cite{2407.10168,2412.02487}, each incorporating free functions that encapsulate quantum gravity effects.
By specifying these functions, the resulting spacetime structures can be thoroughly analyzed.
Notably, the third type of constraint eliminates the presence of Cauchy horizons, which may imply its stability against perturbations \cite{2412.02487}.

After the proposal of the first two types of static black holes in EQG, extensive studies have been conducted on topics such as the strong cosmic censorship conjecture \cite{Lin:2024beb}, quasi-normal mode (QNM) oscillation characteristics \cite{Konoplya:2024lch, Malik:2024nhy, Skvortsova:2024msa}, and the relationship between the greybody factor and the QNMs, which was studied in Ref. \cite{Skvortsova:2024msa}.
At the same time, the accretion disk imaging and strong gravitational lensing effects of these black holes were explored in Refs. \cite{Shu:2024tut,Liu:2024wal,Li:2024afr,Heidari:2024bkm,Wang:2024iwt}.
Additionally, using the Newman-Janis algorithm method \cite{Azreg-Ainou:2014pra}, Refs. \cite{Ban:2024qsa,Li:2024ctu} derived the corresponding rotating black holes from these static black holes and analyzed their shadows.
In our previous work \cite{Liu:2024soc}, we focused on their light rings and shadows; using Event Horizon Telescope (EHT) observational data of the supermassive black hole M87* and the central black hole of the Milky Way, we imposed stringent constraints on the range of their quantum parameters.
Therefore, it is natural to further investigate the light ring and shadow features of the third type of black holes in EQG, which remain relatively unexplored regarding their observational signatures, especially now that the EHT Collaboration has produced black hole images of M87* \cite{APJL875-L1,APJL875-L6} and Sgr A* \cite{APJL930-L12,APJL930-L14}.
Indeed, since these groundbreaking observations, the light ring and black hole shadow have been recognized as promising tools for estimating black hole parameters such as mass, spin, and external fields \cite{PRD79-083004,PRD100-024018,PRD102-024004,JHEP0720054,PRD103-044057,PRD104-044028,PRD106-064058,CQG40-165007}.
Furthermore, black hole shadows offer insights into diverse fundamental issues-ranging from extreme gravitational environments and dark matter to cosmic acceleration, extra dimensions, and potential quantum effects of gravity-these topics have been broadly discussed in Refs. \cite{Wang:2023jop,PRD89-124004,PRD98-084063,PRD100-044055,PRD100-044057,PRD100-024020,JHEP1019269,
EPJC80-790,PRD104-064039,PRD103-064026,PRD103-104033,EPJC81-991,APJ916-116,APJ938-2,CTP74-097401,
EPJC82-835,SCPMA66-110411,JCAP1122006,APJ957-103,APJ958-114,EPJC83-619,2404.12223,JCAP0124059,
APJ954-78,CQG40-174002,PRD109-064027,2307.16748,PRD109-124062,JCAP0524032,2401.17689,2406.00579,
2407.07416,2408.03241,Zheng:2024brm,Ali:2024ssf,AraujoFilho:2024mvz,Hazarika:2024cji,PRD92-104031,
Meng:2023wgi,Malligawad:2024uwi,Ye:2023qks,PRD110-064079,PRD97-064021,JCAP0614043,2005.05992,
2107.00834,2206.08601}.

On the other hand, topology has also garnered considerable interest as an effective mathematical framework to explore black hole properties.
Current topological investigations can be divided into two main facets: one facet focuses on thermodynamic properties, including phase transitions \cite{PRD105-104003,PRD107-044026,JHEP0623115,JHEP1123068,PLB854-138722,2404.02526,JHEP0324138,JHEP0624213} and thermodynamic topological classifications \cite{JHEP0624213,PRL129-191101,PRD107-064023,JHEP0123102,PRD107-024024,PRD107-084002,EPJC83-365,
EPJC83-589,PRD108-084041,PLB856-138919,2404.08243,2409.09333,2409.11666,2409.12747,2411.10102}.
The other facet centers on light rings \cite{PRL119-251102,PRL124-181101,PRD102-064039,PRD103-104031,PRD104-044019,PRD105-024049,PRD105-064070,PRD108-104041,PRD109-064050,AP156-102920,AP162-102994,2405.18798}, extended further to timelike circular orbits \cite{PRD107-064006,JCAP0723049,PRD108-084077,2406.13270}, which may provide additional observational footprints of black holes.
In this paper, we will first study the topological properties of the light rings in the new spacetime proposed in Ref. \cite{2412.02487} and show that these rings are both standard and unstable based on their classification.
Subsequently, we will discuss the shadow of the black hole or compact object described by this solution and use the latest EHT observational data to numerically evaluate its shadow's angular radius and impose constraints on the quantum parameter.
The organization of this paper is outlined as follows.
In Sec. \ref{II}, using the topological method \cite{PRL119-251102,PRL124-181101,PRD102-064039}, we derive the light rings of the third type of static black holes in EQG \cite{2412.02487}.
In Sec. \ref{III}, using the geodesic equation, we derive the orbital equations for photons in this spacetime and estimate the angular radius of the supermassive black holes Sgr A* and M87* within this theoretical framework.
In Sec. \ref{IV}, we discuss the shadow of compact objects with $n>0$ in the outer communication region.
Finally, the paper concludes with our summaries in Sec. \ref{V}.

%%%%%%%%%%%%%%%%%%%%%%%%%%%%%%%%%
\section{Topological properties of light rings}\label{II}
%%%%%%%%%%%%%%%%%%%%%%%%%%%%%%%%%

\begin{figure}[t]
\centering
\includegraphics[width=0.25\textwidth]{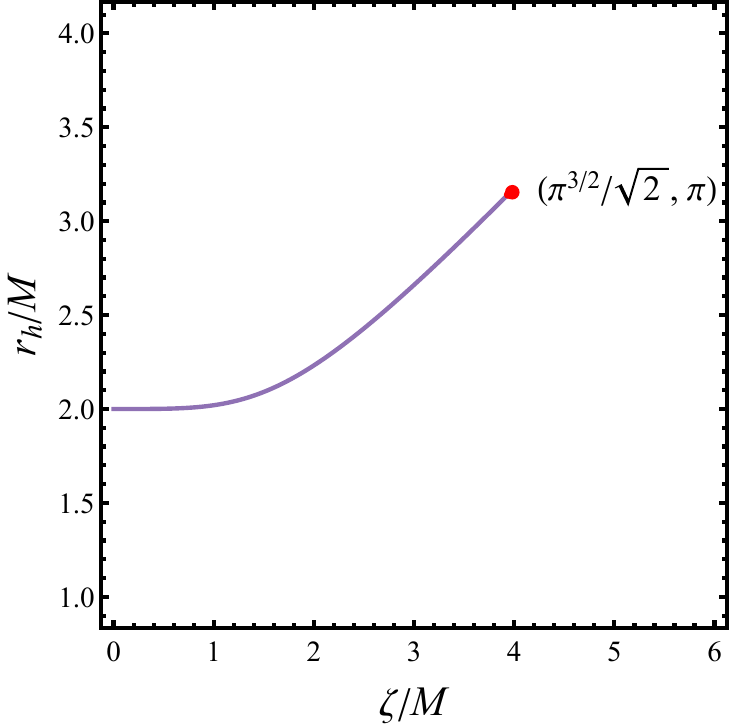}
\caption{The relationship between the quantum parameter and the black hole event horizon is analyzed in a third-type spacetime in EQG, with $n=0$.}
\label{fig1}
\end{figure}

In this section, we first briefly introduce the third type of static solution in EQG, whose metric is given by
\begin{equation}\label{metric3}
ds^2_{(3)}=-f_3^{(n)}dt^2+\frac{1}{\mu_3f_3^{(n)}}dr^2+r^2d\theta^2+r^2\sin\theta d\varphi^2,
\end{equation}
with
\begin{equation}
\begin{aligned}
f^{(n)}_3(r)=&1-(-1)^n\frac{r^2}{\zeta^2}\arcsin\left(\frac{2M\zeta^2}{r^3}\right)-\frac{n\pi r^2}{\zeta^2},\\
\mu_3(r)=&1-\frac{4M^2\zeta^4}{r^6},
\end{aligned}
\end{equation}
in which $M$ is the mass, and $\zeta$, $n$ are the quantum parameters.
Fig. \ref{fig1} illustrates the parameter range within which a compact object, described by the metric $ \meq{metric3} $ with $ n=0 $, can possess an event horizon.
Here, the purple line segment represents the parameter range of a static quantum black hole.
The red dot marks the critical parameter value at which the compact object transitions into a black hole.
When the quantum parameter of the compact object exceeds $ (\pi^{3/2}/\sqrt{2})M $, it no longer possesses an event horizon.
In other words, for a given quantum parameter in a third type of spacetime in EQG, if the mass of the object exceeds $(\sqrt{2}/\pi^{3/2})\zeta$, the object will form an event horizon, thereby classifying it as a black hole.
For $n \neq 0$, if the quantum parameter $\zeta$ is large enough, an external region will exist, and the asymptotic behavior of the solution will closely resemble that of de Sitter space.
On the other hand, if $\zeta$ is small enough, we can consider the conventional cosmological horizon as the event horizon of the black hole, in which case the entire black hole resembles a negative mass object.

Then, we employ the topological method \cite{PRL119-251102,PRL124-181101,PRD102-064039} to study the properties of light rings in the third type of static black holes in EQG with $ n=0 $.
To examine the topological properties of light rings, we begin by introducing a potential function defined as \cite{PRL124-181101,PRD102-064039}
\be
H(r,\theta) = \sqrt{-\frac{g_{tt}}{g_{\varphi\varphi}}} = \frac{\sqrt{f}}{r\sin\theta} \, .
\ee
The radius of the light rings is determined by the condition $\partial_r H = 0$. As outlined in Ref.~\cite{PRD102-064039}, the defining vector $\phi = (\phi^r, \phi^\theta)$ is introduced as
\bea\label{vector}
\phi = \left(\frac{\pp_r H}{\sqrt{g_{rr}}} \, , ~ \frac{\pp_\theta H}{\sqrt{g_{\theta\theta}}}\right) \, .
\eea
The above vector $\phi$ can also be rewrite as
\be\label{v2}
\phi = ||\phi||e^{i\hat{\Theta}} \, ,
\ee
where $||\phi|| = \sqrt{\phi^a\phi^a}$. Notably, the zero point of the vector $\phi$ coincides precisely with the location of the light rings. This indicates that $\phi$ in Eq.~(\ref{v2}) is not well-defined at the light rings. To address this, one can redefine the vector as $\phi = \phi^r + i\phi^\theta$. The normalized vectors are then expressed as
\be
n^a = \frac{\phi^a}{||\phi||} \, , \quad a = 1,2\, ,
\ee
where $\phi^1 = \phi^r$ and $\phi^2 = \phi^\theta$, respectively. Furthermore, a topological current can be constructed using Duan's theory \cite{NPB514-705,PRD61-045004} on $\phi$-mapping topological currents, which is given as
\be\label{jmu}
j^{\mu}=\frac{1}{2\pi}\epsilon^{\mu\nu\rho}\epsilon_{ab}\pp_{\nu}n^{a}\pp_{\rho}n^{b}\, . \qquad
\mu,\nu,\rho=0,1,2,
\ee
where $\pp_{\nu}= \pp/\pp x^{\nu}$ and $x^{\nu}=(t,r,\theta)$. It is easy to check that this topological current satisfies
\be
\pp_{\mu}j^{\mu} = 0 \, .
\ee
This argument clearly shows that $j^\mu$ is nonzero only at the zero points of $\phi^a(x_i)$, that is, when $\phi^a(x_i) = 0$. Consequently, the topological charge in the given parameter region $\Sigma$ can be determined using the following formula:
\be
Q = \int_{\Sigma}j^{0}d^2x = \sum_{i=1}^{N}\beta_{i}\eta_{i} = \sum_{i=1}^{N}w_{i}\, .
\ee
Here, the positive Hopf index $\beta_i$ represents the number of loops made by $\phi^a$ in the vector $\phi$-space as $x^{\mu}$ moves around the zero point $z_i$, the Brouwer degree $\eta_{i}= \mathrm{sign}(J^{0}({\phi}/{x})_{z_i})=\pm 1$. For a closed, smooth loop $C_i$
that encloses the $i$th zero point of $\phi$ while excluding other zero points, the winding number of the vector is then given by
\be
w_i = \frac{1}{2\pi}\oint_{C_i}d\Omega \, ,
\ee
where $\Omega = \arctan(\phi^\theta/\phi^r)$.

%%%%%%%%%%%%%%%%%%%%%%%%%%%%%%%%%%%%%%%%%%%%%%%%%%%%%%%%%%%%%%%%%%%%%%%%%%%%%%
\begin{figure}[t]
\subfigure[The light ring (LR) marked with blue dot is at $(r, \theta) = (3.01,\pi/2)$.]
{\label{fig2a}
\includegraphics[width=0.23\textwidth]{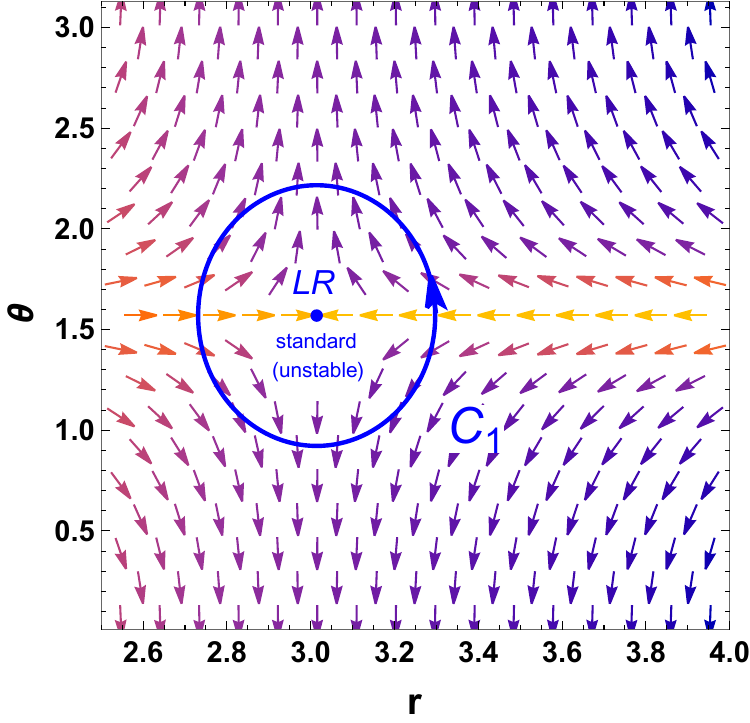}}
\subfigure[The light ring (LR) marked with blue dot is at $(r, \theta) = (3.73,\pi/2)$.]
{\label{fig2b}
\includegraphics[width=0.225\textwidth]{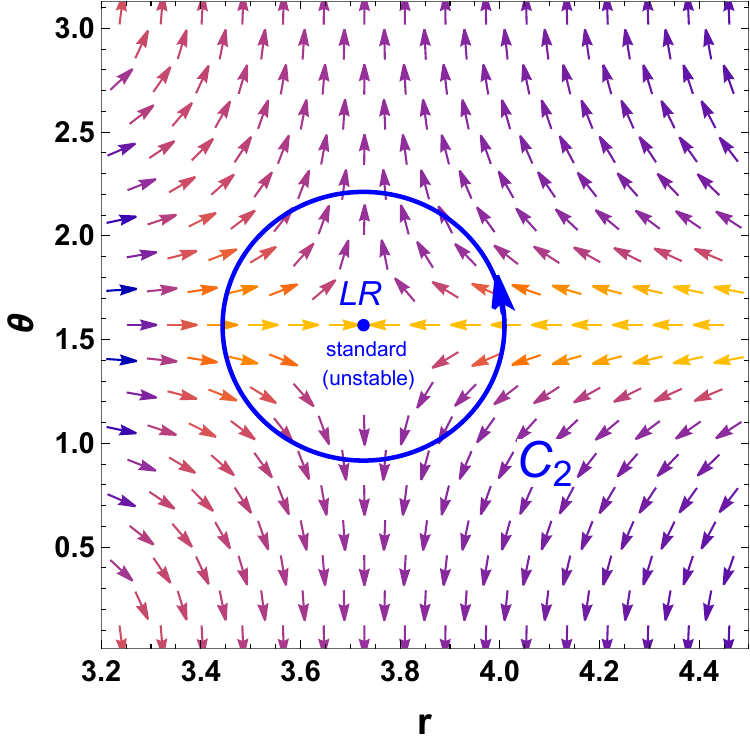}}
\caption{The arrows represent the unit vector field $n$ on a portion of the $r-\Theta$ plane for the this type of static black holes in EQG with (a) $M = 1$, $\zeta = 1$ and (b) $M = 1$, $\zeta = \pi^{3/2}/\sqrt{2}$, respectively. The blue contour $C_i$ are closed loops enclosing the light ring. Obviously, the topological charge of the light ring is always $Q = -1$ in both cases.}
\end{figure}
%%%%%%%%%%%%%%%%%%%%%%%%%%%%%%%%%%%%%%%%%%%%%%%%%%%%%%%%%%%%%%%%%%%%%%%%%%%%%%

In the following, we examine the topological property of light rings for this type of static black hole in EQG.
The unit vector field $n$ is plotted on a portion of the $r\text{-}\theta$ plane in Fig.~\ref{fig2a} with $M = 1$, $\zeta = 1$, and in Fig.~\ref{fig2b} with $M = 1$, $\zeta = \pi^{3/2}/\sqrt{2}$, respectively.
It is evident that a light ring is located at $(r, \theta) = (3.01, \pi/2)$ (i.e., at $r \simeq 3.01M$) in Fig.~\ref{fig2a}, and another light ring is located at $(r, \theta) = (3.73, \pi/2)$ (i.e., at $r \simeq 3.73M$) in Fig.~\ref{fig2b}.
Furthermore, the winding numbers $w_i$ for the blue contours $C_i$ are $w_1 = w_2 = -1$, leading to a total topological charge $Q = -1$ for the light rings of this type of static black hole in EQG.
According to the classification of light rings, these are both standard~\cite{PRL124-181101} and unstable~\cite{PRD102-064039}.
It is worthwhile to delve deeper into the topological light rings of black holes exhibiting unusual horizon structures, such as planar \cite{PRD54-4891}, toroidal \cite{PRD56-3600}, hyperbolic \cite{PRD92-044058}, ultraspinning black holes \cite{PRD103-104020,PRD101-024057,
PRD102-044007,PRD103-044014,JHEP1121031}, and NUT-charged spacetimes \cite{PRD100-101501,
PRD105-124013,2209.01757,2210.17504,2306.00062,2409.06733}, etc.

\section{SHADOWS} \label{III}

\subsection{Photon orbits}\label{Sec.31}
In this section, we briefly overview photon trajectories in third type EQG spacetimes.
Photon motion along geodesics is governed by the Hamilton-Jacobi equation:
\begin{align}\label{EqHJ}
\frac{\pp \mathcal{S}}{\pp \tau}=-\frac{1}{2}g^{ab}\frac{\pp \mathcal{S}}{\pp x^a}\frac{\pp \mathcal{S}}{\pp x^b}.
\end{align}
The affine parameter of the null geodesic is denoted by $ \tau $, while $ \mathcal{S} $ represents the Jacobi action of the photon, which can be separated as:
\begin{align}\label{actionS}
\mathcal{S}=-\mathcal{E}t+L_z\varphi+\mathcal{S}_r(r)+\mathcal{S}_\theta(\theta),
\end{align}
where $ \mathcal{E} $ is the energy, and $ L_z $ is the angular momentum of the photon along the rotation axis.
The terms $ \mathcal{S}_r(r) $ and $ \mathcal{S}_\theta(\theta) $ represent the action's radial and polar components, depending on $ r $ and $ \theta $.
Moreover, the 4-momentum vector of a photon is written as
\begin{align}\label{puu}
p^a=\left(\dot{t},\dot{r},\dot{\theta},\dot{\varphi} \right),
\end{align}
and the metric possesses two obvious Killing vectors, $ \pp_t $ and $ \pp_{\varphi} $, allowing us to compute the conserved quantities:
\begin{equation}\label{gabE}
\begin{aligned}
\mathcal{E}=-p_t,&& L_z=p_{\varphi},
\end{aligned}
\end{equation}
where the components $ p_t $ and $ p_{\varphi} $ are related to the metric as follows:
\begin{align}
p_t = g_{tt}\dot{t}, && p_{\varphi} = g_{\varphi\varphi}\dot{\varphi}.
\end{align}
From the null geodesic equation $ g_{ab}\dot{x}^a\dot{x}^b=0 $, we obtain the relation:
\begin{align}\label{nge}
\frac{p^2_\theta}{r^2}+\frac{p^2_\varphi\csc^2\theta}{r^2}-\frac{p^2_t}{f^{(n)}_3}+\mu_3 f^{(n)}_3p^2_r=0,
\end{align}
with the components $ p_r = g_{rr}\dot{r} $ and $ p_{\theta} = g_{\theta\theta}\dot{\theta} $.
Subsequently, by combining these definitions with the Hamilton-Jacobi equation \meq{EqHJ} and the Carter constant $ \mathcal{K} $ \cite{Carter:1968rr}, which allows the angular and radial parts of the Eq. \meq{nge} to be separated, the equations of motion for all photons are derived as follows:
\begin{equation}
\begin{aligned}\label{taut}
&\dot{t}=\frac{\mathcal{E}\zeta^2}{\zeta^2-i^{2n}r^2 \arcsin\left(2M\zeta^2/r^3\right)-n\pi r^2}, \\
&\dot{r}=\frac{\sqrt{\mathcal{R}(r)}}{r^2},\\
&\dot{\theta}=\frac{\sqrt{\Theta(\theta)}}{r^2},\\
&\dot{\varphi}=-\frac{L_z}{r^2 \sin^2\theta}.
\end{aligned}
\end{equation}
Here, the functions $ \mathcal{R}(r) $ and $ \Theta(\theta) $ defined as:
\begin{align}
&\begin{aligned}
\mathcal{R}(r)=&\mathcal{E}^2\left(r^4-\frac{4M^2\zeta^4}{r^2}\right)-\frac{(\mathcal{K}+L_z^2)(r^6-4M^2\zeta^4)}{r^4\zeta^2}\\
&\times\left[\zeta^2-i^{2n}\arcsin\left(2M\zeta^2/r^3\right)-n\pi r^2\right],\\
\end{aligned}
\end{align}
\begin{align}
\Theta(\theta)=\mathcal{K}-L_z^2\cot^2\theta.
\end{align}
These equations \meq{taut} describe the propagation of light around a compact object of the third type in EQG.

The unstable spherical orbits play a crucial role in determining the boundary of a compact object's shadow.
The spherical orbits satisfy the conditions
\begin{align}
\dot{r}=0,~~~~~ \text{and} ~~~~~ \ddot{r}=0,
\end{align}
which lead to $ \mathcal{R}(r_p)=0 $ and $ \mathcal{R}'(r_p)=0 $.
To further characterize these orbits, one can introduce the impact parameters $ \xi=L_z/\mathcal{E} $ and $ \eta=\mathcal{K}/\mathcal{E}^2 $ \cite{Chandrasekhar}, specifically:
\begin{equation} \label{PSs}
\eta+\xi^2-\frac{r_p^2 \zeta^2}{\zeta^2-i^{2n} r_p^2\arcsin\left(2M\zeta^2/r_p^3\right)-n\pi r_p^2}=0,
\end{equation}
\begin{equation}\label{PSss}
2r_p \zeta^4\sqrt{1-\frac{4M^2\zeta^4}{r_p^6}}-6(-1)^n M\zeta^4=0.
\end{equation}
Here, the resulting impact parameters provide the necessary information, and the real root for $ r_p $ of Eq. \meq{PSss} represents the light ring radius, given by
\begin{align}\label{rppp}
r_p=\sqrt{\left(3+9\chi^{-1/3}+\chi^{1/3}\right)}M,
\end{align}
where
\begin{align}
\chi=27+2\tilde{\zeta}^4+2\sqrt{27\tilde{\zeta}^4+\tilde{\zeta}^8},
\end{align}
is a dimensionless parameter composed of the dimensionless quantum parameter, which is defined as $ \tilde{\zeta}=\zeta/M $.
If $ \tilde{\zeta}=0 $, it corresponds to $ r_p=3M $, which remains consistent with the Schwarzschild black hole case.
We observe that the position of the unstable photon ring is independent of the parameter $n$ in the metric.
It is determined solely by the quantum parameter.
And then, we can obtain the parameter equation at the light ring radius,
\begin{equation}
\begin{aligned}
\eta+\xi^2=&M^2\left[\frac{1}{\Pi}-\frac{n\pi}{\tilde{\zeta}^2}-\frac{i^{2n}}{\tilde{\zeta}^2}\arcsin
\left(\frac{2\tilde{\zeta}^2}{\Pi^{3/2}}\right) \right]^{-1},
\end{aligned}
\end{equation}
where
\begin{align}
\Pi=3+9\chi^{-1/3}+\chi^{1/3}.
\end{align}

When photons originating from a distant source traverse the curved spacetime around a compact object, the intense gravitational field can lens their trajectories toward an observer, while those that cross the event horizon contribute to the formation of a distinct dark region-commonly referred to as the compact object shadow-indicative of the limit beyond which no electromagnetic radiation can escape.
In this paper, we adopt the same orthonormal tetrad as in our previous work on the first and second types of black holes in EQG \cite{Liu:2024soc}.
\begin{equation}\label{zjbj}
\begin{aligned}
e_{(t)}=\left.\frac{1}{\sqrt{-g_{tt}}}\pp_t\right|_{(r_0,\theta_0)},&&
e_{(r)}=\left.-\frac{1}{\sqrt{g_{rr}}}\pp_r\right|_{(r_0,\theta_0)},~~~~\\
e_{(\theta)}=\left.\frac{1}{\sqrt{g_{\theta\theta}}}\pp_\theta\right|_{(r_0,\theta_0)},&&
e_{(\varphi)}=-\left.\frac{1}{\sqrt{g_{\varphi\varphi}}}\pp_\varphi\right|_{(r_0,\theta_0)}.
\end{aligned}
\end{equation}
Here, the observer is located at $ (r_0, \theta_0) $ in the coordinates $ \{t, r, \theta, \varphi\} $, where $ g_{ab} $ represents the metric of a compact object of the third type in EQG.
By expressing the photon's four-momentum through an orthonormal tetrad and relating its spatial components to a three-vector whose magnitude equals the temporal component, one can define the observation angles $ (\alpha,\beta) $ that rigorously quantify the photon's directional distribution within the locally static observer's frame \cite{Liu:2024soc}.

The Cartesian coordinates $ (x,y) $ in the observer's local sky are related to the angular parameters $ (\alpha,\beta) $ as follows:
\begin{equation}
\begin{aligned}
x&=-r_0\tan\beta=-r_0\frac{\sqrt{g_{rr}}p_{\varphi}}{\sqrt{g_{\varphi\varphi}}p_r}, \\
y&=r_0\frac{\tan\alpha}{\cos\beta}=r_0\frac{\sqrt{g_{rr}}p_{\theta}}{\sqrt{g_{\theta\theta}}p_r}.
\end{aligned}
\end{equation}
Under these relations, the compact object's shadow on the observer's celestial plane emerges as the boundary delineated by rays that originate from unstable spherical photon orbits and ultimately plunge into the compact object.

\subsection{Apparent Shape and Its Observation Limitations}\label{Sec41}
For a compact object of the third type in EQG, when $n=0$, the spacetime describes an asymptotically flat geometry.
In this scenario, by placing the observer at an effectively infinite distance and specifying their inclination angle $ \theta_0 $, the observed coordinates $ (x,y) $ on the celestial sphere can be directly related to the constants $ \xi $ and $ \eta $ that characterize the null geodesics forming the compact object's shadow.
Taking the limit $r_0 \to \infty$, we have:
\begin{align}\label{XXX}
x= -\xi \csc \theta_0, &&  y= \sqrt{\eta - \xi^2 \cot^2 \theta_0}.
\end{align}
where $ (x,y) $ represent the apparent position on the observer's sky, delineating the boundary of the compact object's shadow.
It is straightforward to observe that the celestial coordinates satisfy:
\begin{align}\label{X2Y2}
x^2+y^2=\eta+\xi^2=M^2\left[\frac{1}{\Pi}-\frac{1}{\tilde{\zeta}^2}\arcsin
\left(\frac{2\tilde{\zeta}^2}{\Pi^{3/2}}\right) \right]^{-1}.
\end{align}

\begin{figure}[t]
\centering
\includegraphics[width=0.25\textwidth]{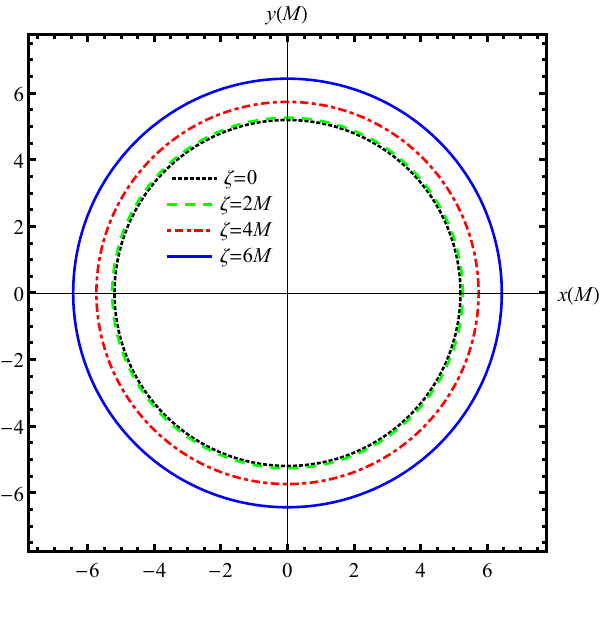}
\caption{Shadow contours under different quantum parameters for the third type of static compact objects in EQG.}
\label{fig3}
\end{figure}
To visualize the impact of the quantum parameter on the shadows of third type black holes in EQG, we have constructed comparative shadow diagrams for different values of this parameter.
As illustrated in Fig. \ref{fig3}, an increase in the quantum parameter results in the expansion of the compact object shadow contours.
This trend sharply contrasts with the findings for the first and second types of static black holes in EQG reported in \cite{Liu:2024soc}.
Specifically, in the first type of static black holes, the quantum parameter causes the shadow contours to shrink, whereas in the second type, the quantum parameter does not affect the shadow contours \cite{Liu:2024soc}.

Currently, we aim to utilize the existing EHT data on the angular radii of the shadows of the supermassive black holes M87* and Sgr A* to constrain the parameters of the third type of compact object within EQG.
For an approximate estimation, employing the metric \meq{metric3}, we calculate the angular radius of the shadow, defined as $ \theta_{\text{BH}} = R_s \frac{\mathcal{M}}{D_O} $, where $D_O$ represents the distance from the observer to the black hole.
Here, according to Eq. \meq{X2Y2}, the compact object shadow radius is
\begin{align}\label{Rss}
R_s=\sqrt{x^2+y^2}=M\left[\frac{1}{\Pi}-\frac{1}{\tilde{\zeta}^2}\arcsin
\left(\frac{2\tilde{\zeta}^2}{\Pi^{3/2}}\right) \right]^{-1/2}.
\end{align}
Specifically, for the approximate estimation of the mass  $ \mathcal{M} $ of the supermassive black hole located at a distance $ D_O $ from the observer, the angular radius $ \theta_\text{BH} $ can be expressed as \cite{Amarilla:2011fx},
\begin{align}
\theta_\text{BH} = \frac{ 9.87098 \times 10^{-6} \mathcal{M}}{\sqrt{\frac{1}{\Pi}-\frac{1}{\tilde{\zeta}^2}\arcsin
\left(\frac{2\tilde{\zeta}^2}{\Pi^{3/2}}\right)}M_\odot}  \left(\frac{1 \text{kpc}}{D_O}\right) \mu\text{as}.
\end{align}
The latest observations give Sgr A* a mass of $ \mathcal{M} = 4.0 \times 10^6 M_\odot $ and a distance of $ D_O = 8.3 $ kpc from the observer \cite{APJL930-L12}.
For M87*, the mass is $ \mathcal{M} = 6.5 \times 10^9 M_\odot $ with an observer distance of $ D_O = 16.8 $ Mpc \cite{APJL875-L6}.

\begin{figure}[t]
\centering
\includegraphics[width=0.25\textwidth]{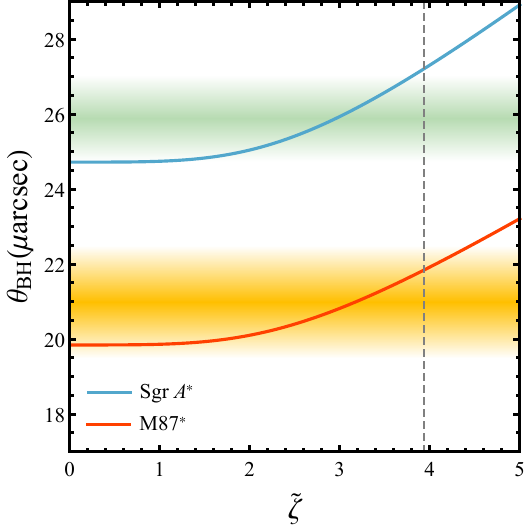}
\caption{The estimation of the angular radius of the supermassive black holes Sgr A* and M87* using the metric of the third type of compact object in EQG.
The black dashed line represents the critical threshold of parameters that determine the existence of the compact object's event horizon.}
\label{fig4}
\end{figure}

Fig. \ref{fig4} displays the calculated angular radius of the supermassive black holes located at the Galactic center (Sgr A*) and in the galaxy M87 (M87*) using a compact object metric within EQG.
In this figure, we employ different gradient colors in the background to broadly highlight the angular radius data for Sgr A* and M87*, aligning them with the general range of recent observational measurements.\footnote{The angular diameters of the black holes Sgr A* and M87* have been measured as $51.8 \pm 2.3 \mu\text{as}$ \cite{APJL930-L12} and $42 \pm 3.0 \mu\text{as}$ \cite{APJL875-L1}, respectively.}
Specifically, the gradient green and gradient yellow regions precisely correspond to the observational ranges of Sgr A* and M87*, respectively, with the fading of color indicating that the observational error margins have been exceeded.
Furthermore, we introduce a reference line at $ \tilde{\zeta}=\sqrt{\pi^3/2} $, which signifies the critical quantum parameter threshold for a compact object to qualify as a black hole.
Interestingly, the range of quantum parameters that classify compact objects as black holes with event horizons is highly consistent with the constraints on quantum parameters imposed by current black hole shadow observations.
This consistency supports the classification of the third type of compact object in EQG as black holes possessing event horizons.

\subsection{Lensing rings}\label{Sec43}
In our previous work \cite{Liu:2024soc}, we investigated the first and second types of static spacetimes in EQG \cite{2407.10168}, demonstrating that for the first type, both the lensing ring and the shadow contours decrease as the quantum parameter $ \zeta $ increases.
In contrast, for the second type, the shadow remains unaffected by the quantum parameter, whereas the lensing ring shrinks with increasing $ \zeta $.
These findings naturally lead us to consider how the third type of spacetime in EQG will behave under similar conditions.
To explore this, we employ the numerical backward ray-tracing method \cite{PRD103-044057} to examine the shadows cast by compact objects.
The specific numerical implementation for EQG spacetimes can be found in \cite{Liu:2024soc}, and since this paper serves as a follow-up to that work, we will not repeat the details here.
By visually representing the portion of the shadow contained within the lensing ring, this method provides a potential criterion for distinguishing various compact objects in EQG.

\begin{figure}[t]
\centering
\includegraphics[width=0.45\linewidth]{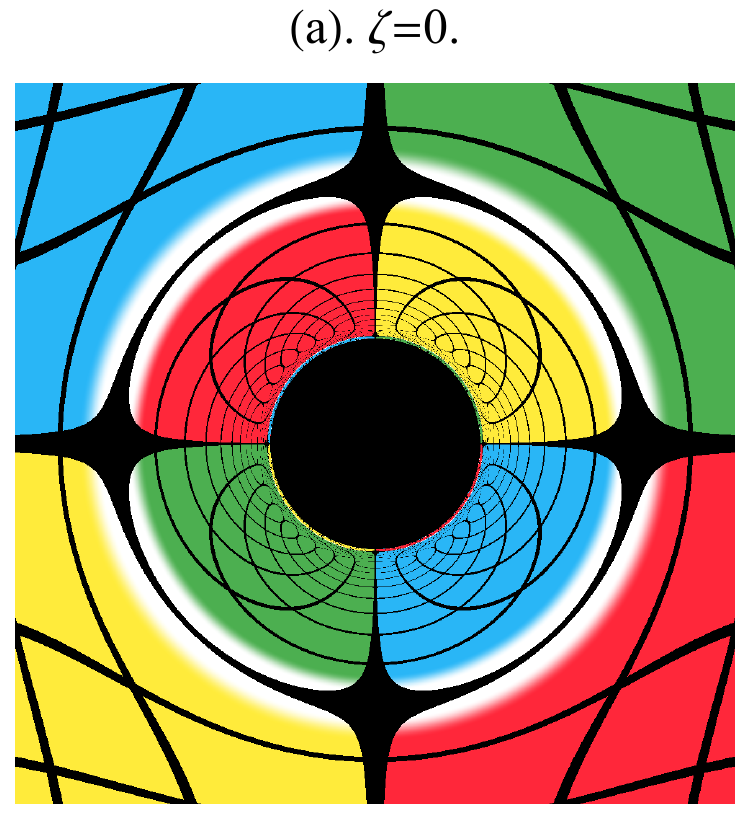}
\includegraphics[width=0.45\linewidth]{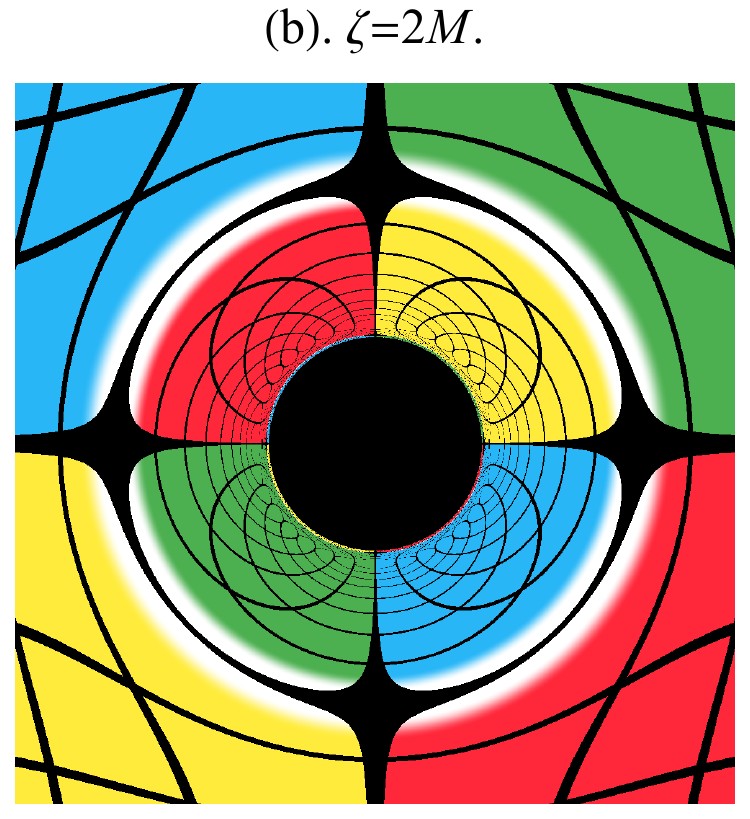}
\includegraphics[width=0.45\linewidth]{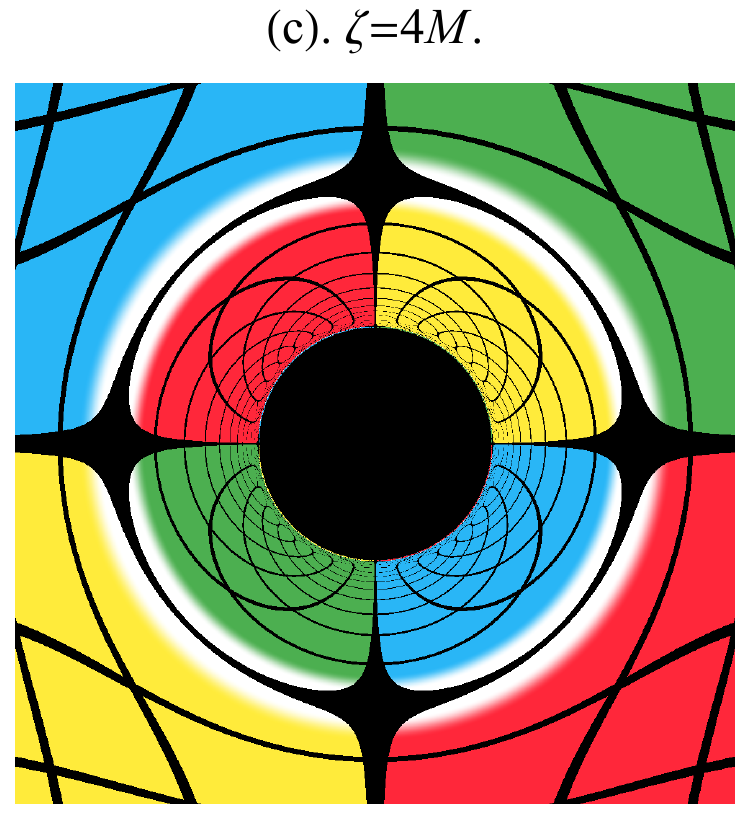}
\includegraphics[width=0.45\linewidth]{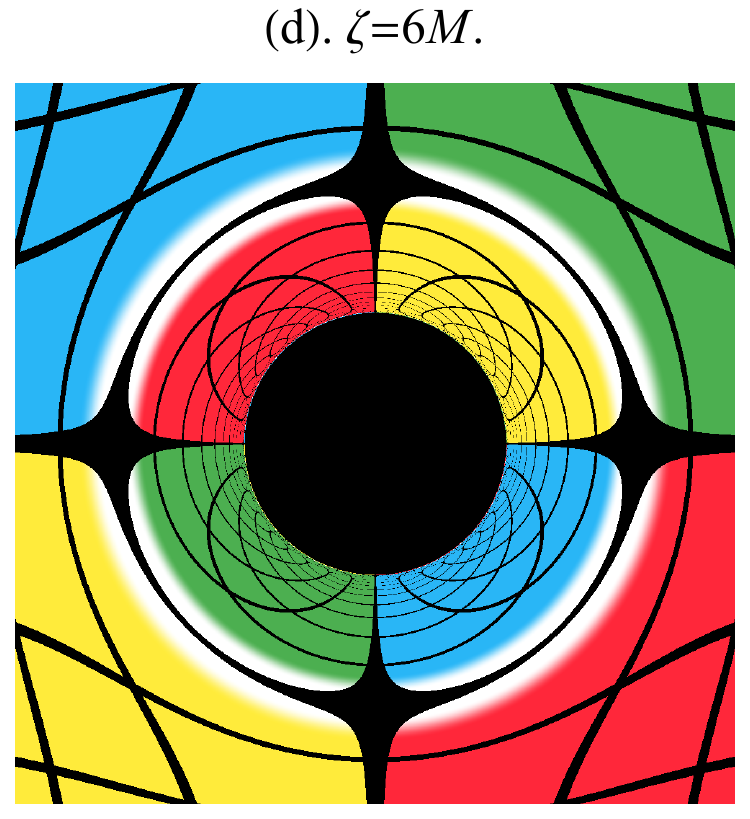}
\caption{The shadows and lensing rings of the third type of static compact object in EQG, characterized by a quantum parameter $\zeta$ for the case $n=0$, as seen by an observer at $\theta_0 = \pi/2$.}
\label{fig7}
\end{figure}
Now, we discuss the impact of the quantum parameter $\zeta$ on the third type of static compact object shadow and lensing rings.
As illustrated in Figs. \ref{fig7}, the shadow radius of the third-type compact object increases as the quantum parameter $ \zeta $ grows.
Comparing Fig. \ref{fig7}(a) and Fig. \ref{fig7}(b), the difference in radius is quite small, indicating a slow, gradual change.
In contrast, when moving from Fig. \ref{fig7}(b) to Fig. \ref{fig7}(c), or from Fig. \ref{fig7}(c) to Fig. \ref{fig7}(d), the radius expansion is noticeably more pronounced.
This observed pattern aligns well with the analytical predictions presented in the previous subsection.
However, the white halo, which delineates the lensing ring, remains invariant under increasing quantum parameters.
Consequently, as the shadow region enlarges, it occupies a progressively greater portion of the lensing ring.
Although this trend is the complete opposite of what is observed for the second type of black holes in EQG, it nonetheless produces a parallel qualitative outcome: as the quantum parameter grows, the shadow progressively dominates the lensing ring's apparent area.
As a result, the combination of a stable lensing ring and an expanding shadow boundary suggests that these third type objects offer a distinct signature that could be probed observationally, potentially serving as an indirect marker for underlying quantum corrections to the gravitational field.

\section{Other Discussions for $ n\neq0 $}\label{IV}
For a third-type compact object in EQG, when $n > 0$, the spacetime structure closely resembles the asymptotically de Sitter (dS) geometry.
In this case, the quantum parameter $ \zeta $ and $n$ together play a role analogous to the cosmological constant in GR.
We can consider a comparison between the Schwarzschild-de Sitter (SdS) spacetime in GR and the third-type spacetime in EQG.
We assume the following relation between the effective cosmological constant and the quantum parameter: $\zeta = \sqrt{3 n \pi / \Lambda}$.
This reparametrization does not imply any fundamental theoretical constraint or physical correlation between $\Lambda$ and $ \zeta $ or $n$; rather, it is employed solely to simplify the form of the third term in the metric and thus facilitate the examination of the spacetime's asymptotic behavior.
As a result, the metric function can be rewritten as:
\begin{equation}\label{fff}
f_3^{(n)}(r)=1-\frac{i^{2n}r^2}{\zeta^2}\arcsin\left(\frac{2M\zeta^2}{r^3}\right)-\frac{\Lambda(\zeta)}{3}r^2.
\end{equation}
For the above equation, whether we consider an expansion at infinity or an expansion with respect to the quantum parameter $ \zeta $, we can always obtain the following approximation:
\begin{equation}
f_3^{n}(r)\sim 1-i^{2n}\frac{2M}{r}-\frac{\Lambda(\zeta)}{3}r^2,
\end{equation}
where, for even integer values of $ n $ (i.e., $ n=2,4,6,... $), the function corresponds to the metric function of the SdS spacetime.
Conversely, for odd integer values of $ n $ (i.e., $ n=1,3,5,... $), its gravitational effects resemble those of a negative mass object \cite{2412.02487}.
Asymptotically dS geometries typically feature a cosmological horizon, defined as the largest root of the metric function $f(r)=0$.
By comparing the locations of these horizon-defining roots in third-type EQG spacetime with those in the standard SdS solution, we can gain a preliminary understanding of how the two spacetimes resemble or differ in their asymptotic structure.
To ensure the existence of a domain of outer communication, the effective cosmological constant must be sufficiently small, which in turn implies a sufficiently large quantum parameter.
Under these conditions, the SdS spacetime possesses an event horizon, whereas the third-type compact object in EQG does not.
The formation of an event horizon for the third-type object would require a smaller quantum parameter (per unit mass), a scenario that precludes the existence of a domain of outer communication.
Since our analysis focuses on regimes where such a domain exists, we do not consider cases with too small a quantum parameter in this work with $ n>0 $.

\begin{figure}[t]
\centering
\includegraphics[width=0.25\textwidth]{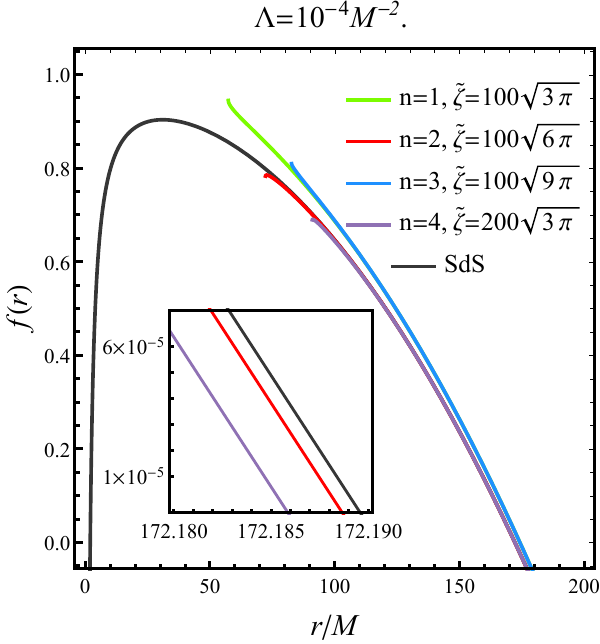}
\caption{The figure illustrates the positions of cosmological horizons under different quantum parameters with the same effective cosmological constant; they are distinct.}
\label{fig5}
\end{figure}

Fig. \ref{fig5} illustrates the behavior of the metric function under the same cosmological constant.
We find that for even integer values of $n$ (specifically, $n=2$ and $n=4$), the metric function's asymptotic behavior closely resembles that of the SdS spacetime.
In these cases, the position of the cosmological horizon decreases as $n$ increases.
Conversely, for odd integer values of $n$ (namely, $n=1$ and $n=3$), the results deviate from the SdS spacetime.
Specifically, the position of the cosmological horizon increases with increasing $n$, which is the opposite trend observed for even values of $n$.
This indicates that the quantum parameter associated with third-type compact objects in EQG continues to influence the asymptotic form of the metric, even for even integer values of $n$.
Such influences may manifest in observable properties, such as black hole shadows or QNMs \cite{Berti:2009kk,Burgess:2023pny,Zhang:2024svj,EPJC83-83,Liu:2024oas,Liu:2023uft,Liu:2024oeq,AraujoFilho:2024rcr,
DeFelice:2024eoj,Heidari:2023yjd}, potentially allowing for their detection through astrophysical observations.
Furthermore, this implies that we can employ a method similar to that used for the SdS spacetime to investigate the shadows in scenarios with $n > 0$.
It is worth noting that, as demonstrated by Eq. \meq{rppp} in the previous section, the location of the unstable light ring does not depend on the parameter $n$.
Consequently, the topological properties of the light ring remain unchanged with varying $n$, allowing us to focus on other aspects of the spacetime geometry.
Therefore, we will not revisit the topological properties of the light ring in this section.

For photons, within the domain of outer communication ($r < r_c$), there exist unstable spherical null geodesics.
Similarly to the known unstable circular photon orbits in Schwarzschild spacetime, these geodesics also exhibit instability and act as limiting trajectories for null rays.
More specifically, when one traces past-oriented light rays observed at a given position backward in time, these rays asymptotically approach the unstable spherical lightlike geodesics.
Hence, the photon region determines the boundary curve of the celestial object shadow.
Due to the properties of null geodesics, the direction of a given photon (its tangent vector) is uniquely determined by its initial conditions near the compact object.
By integrating the geodesic equations from the vicinity of the compact object to the observer's location, we can relate the photon's incoming direction at the observer to its initial direction.
For each light ray $\lambda(s)$ with coordinates $t(s)$, $r(s)$, $\theta(s)$, $\varphi(s)$, the tangent vector in a general coordinate system is given by
\begin{equation}\label{lambda1}
\dot{\lambda}=\dot{t}\pp_t+\dot{r}\pp_r+\dot{\theta}\pp_\theta+\dot{\varphi}\pp_\varphi.
\end{equation}
Following the approach outlined in Ref. \cite{2406.00579}, and still employing Eq. \eqref{zjbj}, we can express the tangent vector in the observer's local orthonormal frame as
\begin{equation}\label{lambda2}
\dot{\lambda}=\gamma[\sin\Theta\cos\Psi e_{(\theta)}+\sin\Theta\sin\Psi e_{(\varphi)}+\cos\Theta e_{(r)}-e_{(t)}],
\end{equation}
where $\gamma$ is a scalar factor.
By substituting Eqs. \meq{taut} and \meq{lambda1} into the above expression, we obtain
\begin{align}
\gamma=-\mathcal{E}\zeta\sqrt{\left(\zeta^2-i^{2n}r^2\arcsin\left(\frac{2M\zeta^2}{r^3}\right)-n\pi r^2\right)^{-1}}.
\end{align}
Substituting this back into equation \meq{lambda2}, we find
\begin{align}\label{sintheta}
\sin\Theta=&
\sqrt{\frac{\frac{M^2}{r^2}-\frac{n\pi}{\tilde{\zeta}^2}-\frac{i^{2n}}{\tilde{\zeta}^2}\arcsin\left(\frac{2M\zeta^2}{r^3}\right)}
{\frac{1}{\Pi}-\frac{n\pi}{\tilde{\zeta}^2}-\frac{i^{2n}}{\tilde{\zeta}^2}\arcsin\left(\frac{2\tilde{\zeta}^2}{\Pi^{3/2}}\right)}}
\Bigg|_{r=r_0}, \\ \label{sinpsi}
 \sin\Psi= & \frac{\xi \csc\theta}{M\sqrt{
\left[\frac{1}{\Pi}-\frac{n\pi}{\tilde{\zeta}^2}-\frac{i^{2n}}{\tilde{\zeta}^2}\arcsin\left(\frac{2\tilde{\zeta}^2}{\Pi^{3/2}}\right)\right]^{-1}}}  \Bigg|_{\theta=\theta_0},
\end{align}
where we assume that the observer is positioned at $ (r_0,\theta_0) $, satisfy $ r_0<r_c $.
Thus, once the trajectory of the photon is determined, we can directly infer the photon's angle of incidence ($\Theta,\Psi$) at the observer.
In order to use analytical parameter to roperesent the boundary curve of the shadow \cite{2406.00579}, we utilized the stereographic projection from the celestial sphere onto a plane, where standard Cartesian coordinates are employed,
\begin{align}
x=-2\tan\left[\frac{\Theta}{2}\right]\sin\left[\Psi\right],\\
y=-2\tan\left[\frac{\Theta}{2}\right]\cos\left[\Psi\right].
\end{align}
This is a set of parameter equations concerning the impact parameter $ \xi $.
Considering that the object under study is a non-rotating celestial body with a standard circular shadow, we can rewrite equation \meq{Rss} using the above expression to obtain:
\begin{align}\label{Rn}
\tilde{R}_s^{(n)}=\sqrt{x^2+y^2}=2\tan\left(\frac{\Theta}{2}\right).
\end{align}
Now, by simultaneously solving Equations \meq{sintheta} and \meq{Rn}, and specifying the mass of the compact object as well as its quantum parameter, we can determine the shadow radius at the position $r_0$.
It is important to note that this method produces results that depend on the observer's location, as the location influences the observation angle $ \Theta $.
Therefore, $ \tilde{R}^{(n)}_s $ is only applicable for analyzing the behavior of the shadow radius with respect to the parameter and does not represent the intrinsic size of the shadow radius at this parameter.

\begin{figure}[t]
\centering
\includegraphics[width=0.25\textwidth]{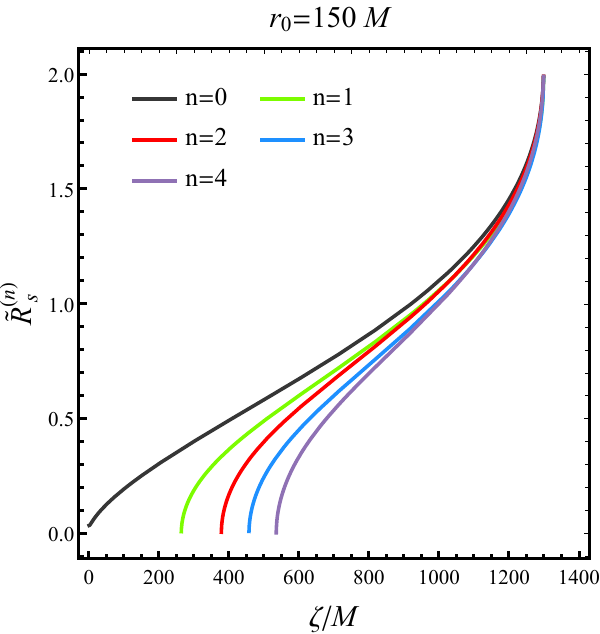}
\caption{The variation trend of the relative shadow radius with respect to the quantum parameter $\zeta$ under different values of $n$.}
\label{fig6}
\end{figure}
Fig. \ref{fig6} illustrates the predicted relative shadow radius with respect to the quantum parameter $\zeta$ for a third-type compact object in EQG as seen by an observer located at $r_0 = 150 M$.
The black line segment corresponds to the case of $n=0$, starting from $\zeta=0$.
In contrast, the other line segments represent scenarios with $n>0$.
For these $n>0$ cases, a domain of outer communication exists at $r_0=150 M$ only when the quantum parameter $\zeta$ is sufficiently large.
As a result, each of these line segments begins at a different $\zeta$ value.
For the first half of the line segments with $n>0$, the shadow radius increases more gradually as $\zeta$ grows, effectively showing a decelerating growth rate.
This behavior differs from the $n=0$ case, where the shadow radius initially grows slowly but then accelerates at relatively small $\zeta$, as depicted in Fig. \ref{fig4}.
For the latter half of the line segments, all $n$ modes exhibit the same trend, reaching the maximum relative shadow radius at a certain quantum parameter value.
This implies that $\Theta = \pi/2$, causing the observer's entire field of view to be dark.
It is necessary to explain that, for the sake of visualization and to facilitate comparisons among different scenarios, we have chosen relatively extreme parameters, leading to theoretically large shadow radii compared to other works that adopt the same calculation method \cite{Eiroa:2017uuq,He:2020dfo,Afrin:2021ggx}.

Based on the data released by the Planck Satellite in 2018 \cite{Planck:2018vyg}, the observed value of the cosmological constant $\Lambda$ is approximately $ \Lambda \sim 1.1056 \times 10^{-52} \, \mathrm{m}^{-2}$, it is not feasible to use current EHT observational data to place meaningful constraints on the parameters for the $n > 0$ cases.
Consequently, the results presented in this subsection are purely theoretical.

\section{Conclusions}\label{V}

Following the introduction of two types of static black holes within EQG, Ref. \cite{2412.02487} has recently derived a third type of spacetime structure, which successfully eliminates the presence of Cauchy horizons.
For the case where the parameter $ n=0 $, their solution asymptotically approaches the Schwarzschild spacetime.
When the quantum parameter $ \zeta=\frac{\pi^{3/2}}{\sqrt{2}}M $ is below the critical value, the solution describes a static quantum black hole with an event horizon.
However, when $ \zeta $ exceeds this critical threshold, the solution represents a static compact object instead of a black hole.

In this paper, we first employ a topological approach to investigate the light rings of the third type of static black holes in EQG \cite{2412.02487}.
Our analysis demonstrates that these light rings are standard and unstable according to established classifications.
Furthermore, we examine the positions of the light rings using the photon trajectory equation, finding results that are consistent with those obtained through the topological method.
Analytical calculations reveal that the locations of the light rings depend solely on the quantum parameter $ \zeta $ and are independent of the parameter $ n $.
For this shadow with $n=0$, its size increases as the quantum parameter grows, while the lensing ring remains unaffected by the quantum parameter.
Although these trends differ from those observed in the two previous types of static black holes in EQG, the combined effect of the shadow and the lensing ring exhibits the same behavior as the second type, namely, as the quantum parameter increases, the shadow occupies a larger portion of the lensing ring.
This difference offers a potential way to distinguish whether the black hole is described by EQG or GR.
It is worth noting that we also performed a numerical evaluation of the angular radius of the shadow of the third type of compact object, using the EHT observation data of M87* and Sgr A*.
The quantum parameter was constrained to $\zeta \lesssim 3.9M \simeq \frac{\pi^{3/2}}{\sqrt{2}}M$, which is highly consistent with the quantum parameter range for classifying compact objects as black holes with event horizons.
This consistency further supports the classification of the third type of compact object in EQG as black holes possessing event horizons.

Finally, we also explored the case of $n > 0$.
Under the same quantum parameter $\zeta$, as $n$ increases, the shadow size decreases.
Additionally, as the quantum parameter increases, the shadow radius for all $n$ modes approaches its maximum value at the same maximum quantum parameter.
We are confident that the results presented in this paper will significantly advance the development of loop quantum gravity theory and contribute to ongoing research aimed at resolving fundamental issues, including black hole singularities and the information paradox.

\acknowledgments
This work is supported by the National Natural Science Foundation of China (NSFC) under Grants No. 12205243, No. 12375053, and No. 12475051;
the Sichuan Science and Technology Program under Grant No. 2023NSFSC1347;
the Doctoral Research Initiation Project of China West Normal University under Grant No. 21E028;
the innovative research group of Hunan Province under Grant No. 2024JJ1006;
and the science and technology innovation Program of Hunan Province under grant No. 2024RC1050;
and Postgraduate Scientific Research Innovation Project of Hunan Province (CX20240531).

\end{document}